\documentclass[showpacs,preprintnumbers]{revtex4-2}
\usepackage{amssymb}
\usepackage{graphicx}
\usepackage{color}
\usepackage{xcolor}

\def\bc{\begin{center}}
\def\ec{\end{center}}

\def\beq{\begin{equation}}
\def\eeq{\end{equation}}

\usepackage{graphicx}
\usepackage{caption}
\usepackage{subcaption}
\usepackage[T1]{fontenc}
\usepackage[utf8]{inputenc}
\usepackage[english]{babel}
\usepackage[hidelinks]{hyperref}
\usepackage{indentfirst}
\usepackage{amssymb}
\usepackage{amsmath}
\usepackage{mathrsfs}
\usepackage{braket}
\usepackage{float}
\usepackage{color}

\begin{document}

\title{Quantum entanglement of multiple excitons in strained graphene}
\author{Gabriel P. Martins$^{1,2,3}$, Oleg L. Berman$^{1,2}$, Godfrey Gumbs$^{2,3}$, and Yurii E. Lozovik$^{4,5}$}

\affiliation{$^{1}$Physics Department, New York City College
of Technology,          The City University of New York, \\
300 Jay Street,  Brooklyn, NY 11201, USA \\
 $^{2}$The Graduate School and University Center, The
City University of New York, \\
365 Fifth Avenue,  New York, NY 10016, USA\\
 $^{3}$Department of Physics and Astronomy, Hunter College of The City University of New York, 695 Park Avenue, \\
New York, NY 10065, USA \\
$^4$ Institute of Spectroscopy, Russian Academy of Sciences, Troitsk, Moscow 142190, Russia \\
$^5$ Research University Higher School of Economics, Moscow 101000, Russia}
\date{\today}

\begin{abstract}

We studied the effects arising from a coherent source of photons on the entanglement between excitons in a strained graphene monolayer. The graphene layer was considered to be embedded in an imperfect optical microcavity. In our investigation, we have studied the entanglement dynamics of systems consisting of up to five excitons, which are treated as atomic-like qubits. Entangled states of multiple qubits are useful in quantum error correction codes. We have monitored the time evolution of the concurrence, three-$\pi$, mutual information, and especially the negativity. We have demonstrated that  coherent pumping can create lasting entanglement between the excitons. However, the entanglement only persists when the rate at which photons are pumped is smaller than the decay rate of the cavity. Our results show that the degree in entanglement between the excitons is increased with the intensity of the strain-induced pseudomagnetic field in the graphene sheet.  Additionally, we have shown that a maximum amount of entanglement occurs at a finite number of excitons in the system which depends on the parameters describing the structure.

\end{abstract}

\maketitle

\section{Introduction}
\label{sec1}

Quantum entanglement is arguably the most interesting feature of quantum mechanics. Ever since the concept was first introduced, it has received ever-increasing attention by researchers \cite{Nielsen_book,Horodecki}. Specifically, in recent years, with the interdisciplinary development of quantum informatics, the general definition, physical properties and measures of quantum entanglement have been comprehensively and deeply studied \cite{conc,tangle,mconc,nega,threepi,mutual}. Due to its non-locality, quantum entanglement has been widely used in quantum information, especially in quantum computing and quantum communication \cite{Penrose,Jozsa,Zou}. 

\medskip
\par

One may describe quantum entanglement as a phenomenon in which two or more systems are coupled together in a way that has no classical counterpart \cite{EPR,Bell}. Performing measurements on one of the entangled entities directly affect all the others by immediately changing their wave functions, even when they are far apart. However, quantum entanglement is  extremely sensitive and even the most subtle interactions with the environment can destroy it \cite{ann,yu,Almeida,Eberly}. For that reason, different approaches to the creation and storage of entangled states that can resist unwanted interactions are the object of many studies \cite{Plenio,Otten,Kauffman,Nourmandipour}.

\medskip
\par
{We note that modern quantum technologies and the prospects for their development are based on the use of two-level quantum systems as qubits. A number of physical realizations of qubits in systems with discrete spectra have been discussed.  The most complex application of  quantum entanglement discussed presently is quantum computation~\cite{Nielsen_book}. Quantum computation heavily uses superposition of multi-qubit systems, i.e., entanglement. Therefore, it is natural that among the most basic paradigms in the field are the production and analysis both of Bell states and of so-called Greenberger, Horne, and Zeilinger  (GHZ) states~\cite{Bruss}, i.e., three-qubit entanglements.

\medskip
\par
In Refs.~\cite{GHZ1,GHZ2}, GHZ proved that for more complicated setups than a two-body experiment, like a three-body experiment, one cannot employ local realistic models to make definite predictions, thereby making quantum  mechanics the only framework where one is able to do that. In their own words: ``$\ldots$ with the appropriate 4-particle (or even 3-particle) system, all one must do is prove that quantum theory holds experimentally, and then we know that it cannot be classically duplicated, $\ldots$'' This means that one needs only to verify the predictions of quantum mechanics for such entangled many-body systems to prove the validity of quantum theory.

\medskip
\par

From the point of view of novel possibilities introduced by quantum computation, it is important to realize that superposition of multi-qubit states makes massive quantum parallelism possible~\cite{Ekert}. It is a fact that a quantum computer can produce result from various different inputs at the same time if the input state is a superposition of the individual information.

\medskip
\par

It is well known that a most important practical problem which a quantum computer faces is decoherence due to coupling to the environment. Somewhat analogous to classical computers, decoherence can be viewed as a quantum analogue of classical noise and it can be overcome using redundant information. Therefore, quantum error correction codes~\cite{Shor,Steane} utilize the possibility of encoding $n$ qubits into the states of a Hilbert-space of higher dimension $N$. One then exploits the additional degrees of freedom to eliminate the quantum noise. Naturally, this implies entangled qubits in a Hilber -space of high dimension.

\medskip
\par

Quantum entanglement is a fundamental property of coherent quantum states and an essential resource for quantum computing~\cite{Nielsen_book}. In large-scale quantum systems, the error accumulation requires concepts for quantum error correction. A first step toward error correction is the creation of genuinely multipartite entanglement, which has served as a performance benchmark for quantum computing platforms such as superconducting circuits~\cite{Dicarlo,Neeley} trapped ions~\cite{Haffner} and nitrogen-vacancy centers in diamond~\cite{Neumann}. Recently, a three-qubit GHZ state has been  generated experimentally using a low-disorder, fully controllable array of three spin qubits in silicon with  a state fidelity of $88.0\%$~\cite{Takeda}. These measurements witness a genuine GHZ class quantum entanglement that cannot be separated into any biseparable state~\cite{Takeda}.

\medskip
\par

The importance of quantum entanglement of three and more qubits for quantum computing is related to the fact that the entangled state of three qubits is characterized by higher fidelity than the entangled state of two qubits. There has been the experimentally observed entangled state of three qubits with a state fidelity of $88.0\%$~\cite{Takeda}. Additionally, the quantum entanglement of three qubits  is very useful for quantum error correction codes.}

\medskip
\par

In recent work, three-qubit states have been generated experimentally using a low-disorder, fully controllable array of three spin qubits in silicon with a state fidelity of $88.0 \%$ \cite{fidelity}. The study of quantum entanglement between multiple qubits is an interesting topic because it has been shown that such states can be employed in quantum error correction codes \cite{Shor,Steane}.

\medskip
\par

Excitons are hydrogen-atom-like quasiparticles consisting of the bound state between an electron in the conduction band and a hole in the valence band \cite{exc}.  In this paper, we will study the dynamics of quantum entanglement between excitons in a gapped graphene monolayer, embedded in an optical microcavity, subject to a pseudomagnetic field (PMF).

\medskip
\par
The pseudomagnetic field can be associated with stretching of graphene \cite{PMF,PMF2}, or with irradiation of graphene with a laser pulse (see, for example, work in Ref. \cite{PMF3}). In our investigation, we assume a physical implementation with a constant pseudomagnetic field associated, for instance, with stretching of graphene. The excitation of qubits may be carried out parametrically, nonadiabatically, by a short laser pulse. To calculate the dynamics of qubit entanglement, we will use the Tavis-Cummings model \cite{TCHam}. Additionally, we must take into account the damping of the excited qubit due to the fact that we are dealing with an open system.

\medskip
\par

The direction of the pseudomagnetic field is perpendicular to the graphene monolayer. There are two valleys in graphene, and the pseudomagnetic fields in different valleys are antiparallel. Therefore, the system remains invariant with respect to the sign of time (in contrast to the real magnetic field).

\medskip
\par
Unlike a real magnetic field, charges are not included in the Hamiltonian of electrons and holes in a pseudomagnetic field. However, the signs in front of the vector potential depend on the signs of the charge and on the direction of the pseudomagnetic field in the valley. The energy spectrum of excitons in strained graphene has been reported in Ref. \cite{BKL}, where it is shown that the exciton's  otherwise continuous energy dispersion becomes a set of discrete and nondispersive Landau levels.

\medskip
\par

If such a graphene sheet is contained within an optical microcavity which is on or near resonance with the energy gap between the exciton ground state and its first excited state, the excitons can be effectively treated as qubits \cite{2qb}. The entanglement dynamics of such a system embedded in a leaky optical microcavity has been studied in Ref. \cite{2qb}. In Ref. \cite{2qb}, it was shown that, for some initial conditions, quantum entanglement was protected, or even enhanced, by the decay of the microcavity, in agreement with the results of Ref.\ \cite{Plenio}, but eventually decayed to zero in other cases. 

\medskip
\par

In this paper, we study the entanglement dynamics of systems of multiple excitons in a strained sheet of graphene, embedded in an optical microcavity. We assume that all these excitons are coupled to a single cavity mode via a Tavis-Cummings Hamiltonian \cite{TCHam}. We have considered a similar system of two excitons in Ref. \cite{2qb}. In this paper, we generalize our approach \cite{2qb} to study a system of three and more excitons. We consider the cavity to be leaky and we study the effects of both coherent and incoherent pumping the cavity with photons to this dynamics. We first study incoherent pumping in a generic setting and show that pumping the cavity by an incoherent source is unable to create any entanglement beween the qubits. In general, the effect due to an incoherent source of cavity photons is to either decrease quantum entanglement between qubits or, at best, leave it unchanged. We show that coherent pumping the cavity with photons can create entanglement. This result is in agreement with the studies reported in  Ref. \cite{albert}. We also study the way in which this entanglement changes when we change some physical parameters describing the system. We perform an in-depth analysis on the time evolution of various entanglement witnesses for the two and three exciton cases and, after that, we study how the negativity of the system changes when we increase the amount of excitons, up to a total of five excitons.  We study how the entanglement dynamics depends on various system parameters, such as the cavity characteristic width, the frequency at which photons are being pumped to the cavity, the intensity of the strain-induced PMF in the graphene sheet and the cavity's volume and dielectric constant.  The framework used by us in this paper can be easily modified for any other system whose dynamics can be approximated by the Tavis-Cummings model.

\medskip
\par


\medskip
\par 
The remainder of this paper is organized as follows: In Sec. \ref{entangle}, we define some of the most used measures of entanglement, namely the concurrence, the negativity, the three-$\pi$, and the mutual information, which will be used by us in the remainder of the paper. In Sec. \ref{totnegsec}, we present a way to estimate the total amount of entanglement contained in a multipartite system in which it can be hard to precise - and quantize - how many of the subsystems are entangled with each other. In Sec. \ref{realization}, we describe the system we consider, namely that of excitons in a strained monolayer of graphene embedded in an optical microcavity. In the rotating wave approximation, this system can be approximated by the Tavis-Cummings Hamiltonian, in which $N$ non-interacting qubits are coupled to a single cavity mode. We present how the dynamics of such a system will occur when spontaneous decay is taken into consideration as well as both coherent and incoherent pumping of cavity photons. Moving on to Sec. ~\ref{incoh}, we perform a general study of the effects due to incoherent pumping of photons in the entanglement between the qubits and show that such forms of pumping are unable to create any entanglement.  Section \ref{results} shows numerical results from our simulation for our system of interest, and show that, under the right conditions, long-lasting entanglement is created between the excitons. Our results are separated in three subsections. The first deals with systems of only two excitons, the second with systems of three excitons, and the last compares the final value of the entanglement when we increase the number of excitons in the system from two to five. In Sec. \ref{limitations}, the limitations of our model and interpretation of our results are presented. Sec. \ref{conclusion} concludes this work.

\section{Measures of Entanglement \label{entangle}}

In this section, we will define all quantities we will employ when analyzing the quantum entanglement present in the studied multi-qubit systems. So far, no quantity has been proposed that is universally sufficient and necessary for quantum entanglement between more than two qubits. Because of this, many different forms to monitor entanglement currently exist. 

\subsection{Concurrence}

The concurrence was first defined as a measure for the entanglement between two qubits \cite{conc}. The concurrence is an entanglement monotone, meaning that it cannot be created via local operations and classical communication (LOCC). When dealing with two qubits systems, a non-zero entanglement is a sufficient and necessary condition for quantum entanglement, making it the perfect entanglement measure for such systems. 

The concurrence $C$ between 2 qubits is defined as 
\begin{equation}
C = \texttt{Max}\{0,\lambda_1-\lambda_2-\lambda_3-\lambda_4\}, \label{conc}
\end{equation}
where the $\lambda_i$ are the eigenvalues  of $\tilde{\rho}\rho$ in descending order, where $\tilde{\rho}$ is the complex conjugate of the density matrix $\rho$.

Concurrence can be extended to larger systems, but not in a unique way \cite{mconc}. Also, for larger systems, having a non-zero concurrence is no longer a necessary condition for quantum entanglement, although it is still sufficient.

\subsection{Negativity}

Negativity is another quantity used to witness and measure entanglement between subsystems in a larger system. The negativity $\mathcal{N}_{A(B)}$ of subsystem $A$ with respect to subsystem $B$ is defined as the sum of the absolute value of all the negative eigenvalues of the partially transposed density matrix of the total system, where the partial transpose is taken over subsystem $A$ \cite{nega}. Mathematically,
\begin{equation}
\mathcal{N}_{A(B)} = \dfrac{\texttt{Tr}\sqrt{{\rho^{T_A}}^2}-1}{2},\label{negdef}
\end{equation}
where the partial transpose $\rho^{T_A}$ of density matrix $\rho$ is defined by the relation
\begin{equation}
\rho^{T_A}_{AB,A^\prime B^\prime} = \rho_{A^\prime B,A B^\prime}. 
\end{equation}

The negativity is extremely relevant, since it can be applied to any composite system, regardless of the number of subcomponents or the dimensionality of the Hilbert space of it's subcomponents. A non-zero value of negativity is a condition sufficient, but not necessary, for quantum entanglement.

\subsection{Three-$\pi$}

The Three-$\pi$ is an entanglement witness aimed at measuring only three-way entanglement between 3 qubits, discarding pairwise entanglement. The Three-$\pi$, $\pi_A$ related to qubit $A$ on a set of three qubits is defined as \cite{threepi}
\begin{equation}
\pi^A = {\mathcal{N}_{A(BC)}}^2-{\mathcal{N}_{A(B)}}^2-{\mathcal{N}_{A(C)}}^2,
\end{equation}
where $\mathcal{N}_{A(BC)}$ is the negativity of the total system with respect to qubit $A$, and $\mathcal{N}_{A(B)}$ ($\mathcal{N}_{A(C)}$) is the negativity with respect to qubot $A$ when we trace out subsystem $C$ ($B$). Since the Three-$\pi$ is qubit dependent, one good way to measure 3-way entanglement in the total system, is the average of the Three-$\pi$'s of all 3 qubits,
\begin{equation}
\pi_{ABC}=\dfrac{1}{3}(\pi_A+\pi_B+\pi_C).\label{threepidef}
\end{equation} 
Since three-$\pi$ is derived from the negativity, a non-zero value of it is also a sufficient albeit not necessary condition for quantum entanglement.

\subsection{Mutual Information}

The mutual information $MI_{AB}$ between subsystems $A$ and $B$ is defined as the difference between the sum of the Shannon entropies $\mathcal{S}_A$ and $\mathcal{S}_B$, of subsystems $A$ and $B$, and the total entropy of the combined system $\mathcal{S}_{AB}$ \cite{mutual}, where the Shannon entropy $\mathcal{S}_O$ of system $O$ is given by
\begin{equation}
\mathcal{S}_O = \texttt{Tr}\left(\rho_O\log\rho_O\right),
\end{equation}
where $\rho_O$ is the density matrix of system $O$. The mutual information $MI_{AB}$ is, therefore, given by
\begin{equation}
MI_{AB} = \texttt{Tr}\left(\rho_A\log\rho_A\right) + \texttt{Tr}\left(\rho_B\log\rho_B\right) -\texttt{Tr}\left(\rho_{AB}\log\rho_{AB}\right),\label{MIdef}
\end{equation}
where $\rho_{AB}$ is the density matrix of the combined system $AB$ and $\rho_A = \texttt{Tr}_A\rho_{AB}$ ($\rho_B = \texttt{Tr}_B \rho_{AB}$) is the reduced density matrix of system $A$ ($B$) alone.

For pure states, having non-zero mutual information, is a necessary and sufficient condition for entanglement. However, a classical mixture of separable states can have non-zero mutual information even though it shows no quantum entanglement whatsoever, although, even for mixed states, a non-zero value of mutual information is still a necessary condition for quantum entanglement. Mutual information is also not an entanglement monotone, meaning that it's value can increase via LOCCs.

\section{An Estimate on Total Entanglement within a multi-qubit system}\label{totnegsec}

Every entanglement witness presented here was created with the intention of quantifying the entanglement between one subsystem and all other subsystems within a bigger system. This, however, is not always a good indicative of the total amount of quantum entanglement shared between all the subsystems. In this section we will provide a way to estimate the total amount of entanglement within a multi-qubit system based on the negativity, which we name \textit{total negativity}. We will provide clear values for the upper and lower bounds of the total negativity. 

\medskip
\par 
This is evident in the trivial cases: for example, let us consider a 3 qubit system in state $\ket{\psi} = \ket{0}_1 \otimes \ket{\Psi_+}_{2,3}$, where $\ket{\Psi_+} = \dfrac{1}{\sqrt{2}}\left(\ket{01} + \ket{10}\right)$ is one off the maximally entangled Bell states. In this system, qubit number 1 is not entangled to the other qubits and any entanglement witness measured on qubit 1 will be unable to observe any entanglement, even though quantum entanglement is clearly present in the composed system. 

\medskip
\par 
One naive attempt to quantify the total amount of entanglement in a multi-qubit system would be to sum over the individual negativities with respect to each of it's subsystems. Doing that, however, is not a good approach. This is due to the fact that quantum entanglement is always shared by at least two subsystems. When we calculate how entangled is, for example, subsystem $A$ to subsystems $B$ and $C$ and then calculate the analogous for subsystem $B$, we are counting the partial entanglement shared only by subsystems $A$ and $B$ twice. One of the most extreme examples of how this approach is erroneous, is the $N$ qubit Greenberger-Horne-Zeilinger (GHZ) state \cite{GHZ1,GHZ2}, $\ket{GHZ} = \dfrac{1}{\sqrt{2}}\left(\ket{0_1\hdots 0_{N}} + \ket{1_1\hdots 1_N}\right)$. In this particular state, entanglement can only be seen when analyzing all qubits together. This is evidenced by the fact that disregarding any of the qubits - mathematically, tracing any of the qubits out - leaves the reduced system in a state $\rho_{\mathrm{red}}$ which is a mixture of completely separable states, $\rho_{\mathrm{red}} = \dfrac{1}{2}\ket{0_1\hdots 0_{N-1}}\bra{0_1\hdots 0_{N-1}} + \dfrac{1}{2}\ket{1_1\hdots 1_{N-1}}\bra{1_1\hdots 1_{N-1}}$.  This means that, if even a single qubit is for any reason not accessible, the remaining system for the $N-1$ qubits is a classical statistical mixture in which all qubits are on the state $\ket{0}$ with probability 0.5, or all qubits are on the state $\ket{1}$, also with probability 0.5. The absence of crossed therms in this reduced density matrix means that no measure on the resulting $N-1$ qubit system can detect non-trivial correlations due to quantum entanglement. For that reason, when all $N$ qubits are accessible, we can say that the entanglement observed by \textit{any} of the qubits, $\mathcal{N}_i$  is actually the total entanglement in the combined system $\mathcal{N}_{tot}$ which means that, for the particular cases in which entanglement can only be observed when all qubits are accessible, summing over the entanglement experienced by individual qubits wields $\sum_i^N \mathcal{N}_i = \sum_i^N \mathcal{N}_{tot} = N \times \mathcal{N}_{tot}$. Which means that $\mathcal{N}_{tot} = \dfrac{1}{N}\sum_i^N \mathcal{N}_i$. 

\medskip
\par 
 
Similarly, if, for example, we know that in our composite system of $N$ subsystems, every single form of entanglement is always shared by sets of $k$ subsystems, where $k \leq N$ is an integer number, the total negativity $\mathcal{N}_{tot}$ is going to be equal to
\begin{equation}
\mathcal{N}_{tot} = \dfrac{1}{k}\sum_i^N \mathcal{N}_i, \label{defp1}
\end{equation}
where $\mathcal{N}_i$ is the negativity with respect to susbsytem $i$.  Eq. (\ref{defp1}) can be more easily understood when we analyze a comcrete case. Let us consider, for example, the 6 qubits system on the state $\ket{\psi} = \ket{GHZ}_{1,2,3}\otimes\ket{GHZ}_{4,5,6}$, where $\ket{GHZ}_{1,2,3}$ ($\ket{GHZ}_{4,5,6}$) is the 3 qubit GHZ state for qubits 1, 2, and 3 (4, 5, and 6). In this case, we know all entanglement is shared always by sets of $k = $ 3 qubits in this $N = $ 6 components subystsem. Summing over the entanglements experienced by qubits 1, 2, and 3, overcounts that portion of the entanglement by a factor of three. The same is true for qubits 4, 5, and 6. It is clear that, in this case, we can counteract this overcounting of entanglement by dividing the overall result by a factor of $k=3$, and the total negativity in this composite system will be $\mathcal{N}_{tot} = \dfrac{1}{3}\sum_{i=1}^6 \mathcal{N}_i$.

 Unfortunately, when we calculate the negativity $\mathcal{N}_i$ of subsystem $i$ with respect to the composite system, there is no way in general to properly know how much of it comes from pairwise entanglemenr, how much comes form entanglement shared between three subsystems, four subsystems, and so on. So, apart from very specific cases, evaluating the actual value of $\mathcal{N}_{tot}$ is impossible. However, we can provide clear upper and lower bounds for it. Since the entanglement is shared by at least two subsystems, and at most all the subsystems, we can state that the total negativity always obeys

\begin{equation}
\dfrac{1}{N}\sum_i^N \mathcal{N}_i \leq \mathcal{N}_{tot} \leq \dfrac{1}{2}\sum_i^N \mathcal{N}_i. \label{Ntot}
\end{equation}
Note that the upper and lower bounds of Eq. (\ref{Ntot}), are nothing more than the upper and lower bounds of Eq. (\ref{defp1}), which happens when $k=2$, and $k = N$, respectively.


\section{Excitons in Strained Graphene\label{realization}}

\begin{figure}[H]
\begin{center}
\includegraphics[width=0.5\textwidth]{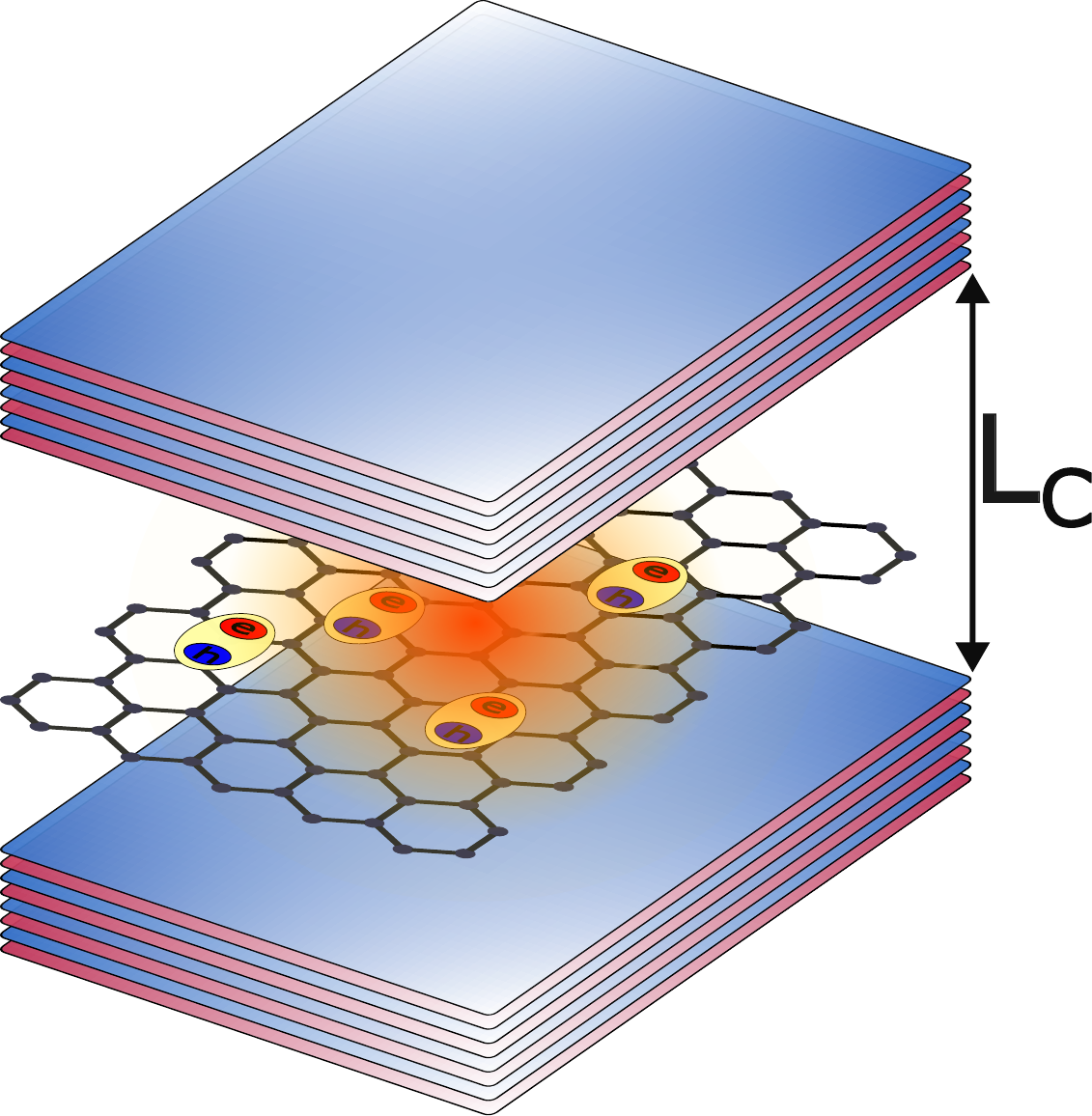}
\end{center}
\caption{(Color online) Schamatics of our considered system. Excitons on a strained sheet of graphene inside an optical microcavity. The microcavity is made of two sets of Bragg mirrors, separated by a distance $L_C$. \label{schem}}
\end{figure}

In this section, we will provide an in-depth description of our system of interest, namely that of excitons in a strained sheet of graphene monolayer embedded in an optical microcavity, as the one depicted in Fig. \ref{schem}. 

\medskip
\par 
When a graphene sheet is put under strain, electrons and holes in it behave analogously to the case in which they are subject to a magnetic field $B$ \cite{PMF}. The so called pseudomagnetic field (PMF) is proportional to the deformation caused by the strain in the graphene sheet. 

\medskip
\par 
There are two types of excitons in the system under consideration - intervalley excitons (the electron and hole are in different valleys) and intravalley excitons (the electron and hole are in the same valley). 

\medskip
\par 
For intravalley excitons, the electron and hole are subject to a pseudomagnetic field in the same valley. Therefore, their properties in a pseudomagnetic field are the same as for two-dimensional excitons in a real perpendicular magnetic field. In a strong pseudomagnetic field they have the same dependence on the magnetic momentum \cite{Loz1}, and they also form a Bose-Einstein condensate of non-interacting excitons at low temperatures \cite{Loz2,Loz3,Loz4}.   In contrast, for intervalley excitons, oppositely directed pseudomagnetic fields act on the electron and hole, and as a result, their effect on the electron and hole turns out to be the same. So intervalley excitons behave like particles with two identical charges. Therefore, in a strong pseudomagnetic field, a quantum Hall effect is possible for a system of intervalley excitons \cite{BKL}. In this paper, we consider the latter. In such an environment, intervalley excitons have a discrete, dispersionless energy set.

\medskip
\par 
The energy levels of the pseudomagnetic excitons (PME) are given by \cite{BKL}

\begin{equation}
\varepsilon_{n,\tilde{n},\tilde{m}} =  \varepsilon_{0n\tilde{n}} + E^\prime_{\tilde{n},\tilde{m}},
\end{equation}
where $\varepsilon_{0n\tilde{n}}$ are the energy eigenvalues of the non-interacting electron-hole pair, and $E^\prime_{\tilde{n},\tilde{m}}$ are the exciton binding energies. 
For direct excitons (when the electron and hole are on the same graphene monolayer), we have

\begin{eqnarray}
&\varepsilon_{0n\tilde{n}} = \hbar \omega_{cy}\left(\dfrac{1}{2} + n\right) + \hbar \tilde{\omega}_{cy}\left(\dfrac{1}{2} + \tilde{n}\right)& ; \\
E^\prime_{0,0} = - E_0 ; & E^\prime_{0,1} = - \dfrac{E_0}{2} ;& E^\prime_{1,0} = - \dfrac{E_0}{4};\\
&E_0 = \sqrt{\dfrac{\pi}{2}}\dfrac{k e^2}{\varepsilon_d l},&
\end{eqnarray}
where $\omega_{cy} = 2B / M$ (where $M = m_e+m_h$ is the exciton total mass, and $B$ is the PMF caused by strain),    $\tilde{\omega}_{cy} = \tilde{B}/\mu$ (where $\mu = m_e m_h/(m_e+m_h)$ is the exciton reduced mass and $\tilde{B} = (m_e^2 + m_h^2) B/ M^2$), and $l = \sqrt{\hbar/\tilde{B}}$ the pseudomagnetic length. For direct excitons, $m_e = m_h$ which leads to $\omega_c = \tilde{\omega}_c$.

\medskip
\par 
The energy gap $\Delta_{ex}$ between the ground state and first excited state for direct excitons in strained graphene is given by \cite{BKL}
\begin{equation}
\Delta_{ex} = \varepsilon_{0,0,1}-\varepsilon_{0,0,0} =  \sqrt{\dfrac{\pi}{2}}\dfrac{k e^2}{2\varepsilon_d l}. \label{gapstrain}
\end{equation}

\medskip
\par 
The effective mass of electrons and holes in a graphene monolayer is $m_e = \dfrac{\Delta}{v_F^2}$, where $v_F = 10^6$ m/s is the Fermi velocity of electrons in graphene, and $\Delta$ is the energy gap between the valence and conduction bands on graphene. For $\Delta = 0.25$ eV, we have $m_e = 0.044$ $m_0$, with $m_0$ the electron rest mass. 

\medskip
\par 
If such a system is embedded in an optical microcavity which is tuned to be in or near resonance with the energy gap between the ground and first excited state of the PMEs, one can effectively treat the PMEs as qubits \cite{2qb}. In this paper, we consider a set of $N$ such qubits.

\medskip
\par 
 In the limit of low excitonic density, we can disregard exciton-exciton interactions. In this limit, in the rotating wave approximation, this system can be described by the Tavis-Cummigs Hamiltonian \cite{TCHam}
\begin{equation}
\hat{H}_0 = \hbar\omega_c \hat{a}^\dagger\hat{a} + \sum_{j=1}^N \left(\dfrac{\Delta_{ex}}{2} {\sigma_z}_j + g(\hat{a}{\sigma_+}_j + \hat{a}^\dagger{\sigma_-}_j )\right], \label{TCH}
\end{equation} 
where $\hbar\omega_c$ is the energy of the cavity photons, $g$ is the Rabi splitting, $\hat{a}$ is the annihilation operator for photons in the cavity, ${\sigma_z}_j$ is the Pauli matrix for the $j$-th exciton, and ${\sigma_+}_j$ and ${\sigma_-}_j$ are the creation and annihilation operators for excitations in the $j$-th exciton, respectively.  The cavity frequency $\omega_c$ relates to the cavity width $L_C$ as
\begin{equation}
L_C = \dfrac{\pi c}{\sqrt{\varepsilon_d}\omega_c},
\end{equation}
where $\varepsilon_d$ is the dielectric constant of the microcavity. In this system, the Rabi coupling between the excitons and photons is given by \cite{2qb}
\begin{equation}
g = ev_F\left(\dfrac{\pi\hbar^2}{\varepsilon_0\varepsilon_d \Delta_{ex} W}\right)^{\frac{1}{2}}, \label{rabistrain}
\end{equation}
where $W$ is the volume of the microcavity, and $v_F \approx 10^6$ m/s is the Fermi velocity for electrons in graphene.

\medskip
\par 
If the microcavity is leaky, the dynamics of the system will be governed by the master equation \cite{Nielsen_book}
\begin{equation}
\hbar \dot{\rho} = \dfrac{1}{i}\left[ \hat{H}_0,\rho\right]  + \gamma_c\mathcal{L}(\hat{a}) \rho +\gamma_q\sum_{j=1}^N\mathcal{L}({\sigma_-}_j) \rho, \label{master}
\end{equation}
where ${\gamma_c}$ is the cavity decay rate, $\dfrac{\gamma_q}{\hbar}$ is the qubit decay rate,  and the Lindblad superoperators $\mathcal{L}(\hat{O}) \rho$ for the arbitrary operator $\hat{O}$ is defined as
\begin{equation}
\mathcal{L}(\hat{O})\rho = \hat{O}\rho\hat{O}^\dagger - \dfrac{1}{2}\left\lbrace \hat{O}^\dagger\hat{O},\rho \right\rbrace.
\end{equation}  

Recently, it has been shown that in the case in which $\gamma_q\ll\gamma_c$, there exist entangled states that are protected from decaying via cavity emission and, in this limit, some entangled states have a lifetime comparable to $\gamma_q^{-1}$ \cite{2qb}. However, if neither $\gamma_c$ nor $\gamma_q$ can be disregarded in Eq. (\ref{master}), without an external source of pumping, the system will eventually decay to the ground state as $t\rightarrow\infty$, regardless of it's state at $t=0$. 

\medskip
\par 
In this paper, we will analyze the effects of both a coherent and incoherent source of pumping in the cavity mode in the creation and maintenance of entanglement in the system. Incoherent pumping is represented by the addition of a term $\gamma_p\mathcal{L}(\hat{a}^\dagger)\rho$ to Eq. (\ref{master}), where $\gamma_p$ is the pumping rate. Coherent pumping will be treated in the semiclassical approximation, in which the cavity mode is treated quantum mechanically and the pumping mode is treated classically \cite{optics}. The dynamics in this regime is obtained by replacing the Hamiltonian $\hat{H}_0$ in Eq. (\ref{master}) by
\begin{equation}
\hat{H} = \hat{H}_0 + \hbar R_P\left(e^{+i\omega_L t}\hat{a}+e^{-i\omega_L t}\hat{a}^\dagger\right),\label{hpump}
\end{equation}
where $R_P$ is the pumping rate of photons into the cavity by the laser and $\omega_L$ is the laser frequency. Namely, the dynamics of the system will be given by
\begin{equation}
\hbar\dot{\rho} = \dfrac{1}{i}\left[ \hat{H},\rho\right]  + \gamma_c\mathcal{L}(\hat{a}) \rho +\gamma_q\sum_{j=1}^N\mathcal{L}({\sigma_-}_j) \rho, \label{master_pump}
\end{equation}
where $\hat{H}$ from Eq.(\ref{hpump}) took the place of $\hat{H}_0$ in Eq. (\ref{master}).  Throughout our simulations, except when it is clearly stated otherwise, we assume that the laser frequency is in resonance with the frequency of cavity photons, $\omega_L = \omega_c$, as well as the cavity photons are in resonance with the excitonic energy gap $\hbar\omega_c = \Delta_{ex}$. 

\medskip
\par 
PMFs have been experimentally verified up to intensities of about 300 T \cite{Levy}. In our simulations, we consider PMFs of intensities ranging from 10 to 90 T. In this region, the energy gap in the excitons, given by Eq. (\ref{gapstrain}),  goes from $\approx$ 4 meV up to $\approx$ 10 meV. This means that one would need microcavities whose resonant frequencies range from about 5 THz to about 17 THz to perform such an experiment, as well as coherent light sources in the same spectrum. Both cavities and coherent light sources have been achieved in this region \cite{cav,laser}, and, therefore replicating our numerical in a physical setting should be easily achievable.

\section{Incoherent pumping }\label{incoh}

 In this section, we analyze the effects of an incoherent source of cavity photons in the entanglement dynamics between qubits following the Tavis-Cummings Hamiltonian. For that, we numerically solved  for the density matrix $\rho$ the master equation
\begin{equation}
\hbar \dot{\rho} = \dfrac{1}{i}\left[ \hat{H}_0,\rho\right]  + \gamma_c\mathcal{L}(\hat{a}) \rho ++ \gamma_p\mathcal{L}(\hat{a}^\dagger) \rho, \label{masterinc}
\end{equation}
where $\hat{H}_0$ is the Tavis-Cummings Hamiltonian of Eq. (\ref{TCH}). Additionally, $\gamma_c, \gamma_p$ are the cavity decay rate and incoherent pumping rate, respectively.

\medskip
\par 
 In this part, we considered the system starting the dynamics in 4 different states: (a) $\ket{00}$; (b) $\ket{\Psi_-} = \dfrac{1}{\sqrt{2}}\left(\ket{01}-\ket{10}\right)$; (c) $\ket{\Psi_+} = \dfrac{1}{\sqrt{2}}\left(\ket{01}+\ket{10}\right)$; and (d) $\ket{11}$. From our results, we decided not to show the plots corresponding to the system starting at the state $\ket{\Psi_-} = \dfrac{1}{\sqrt{2}}\left(\ket{01}-\ket{10}\right)$, because, just like in the case in which no pumping is present, this state is protected from decay and the overall qubits state remains constant throughout the simulation. No entanglement is created or lost, nor any excitations in the qubits are created in this particular case.

\begin{figure}[H]
\begin{center}
\begin{subfigure}[b]{0.30\textwidth}
\includegraphics[width=\textwidth]{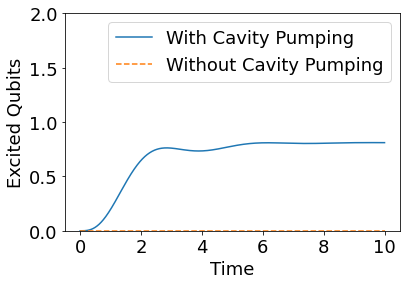}
\caption{}
\end{subfigure}
\begin{subfigure}[b]{0.30\textwidth}
\includegraphics[width=\textwidth]{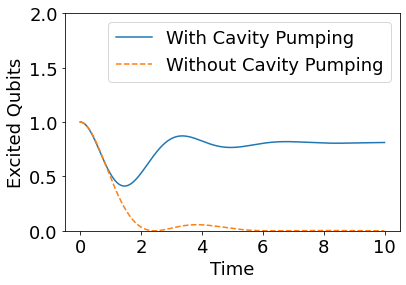}
\caption{}
\end{subfigure}
\begin{subfigure}[b]{0.30\textwidth}
\includegraphics[width=\textwidth]{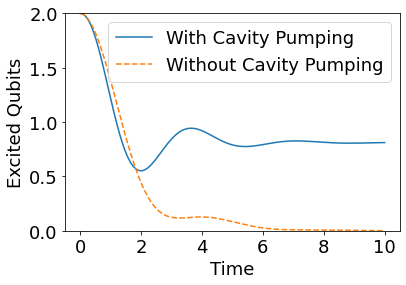}
\caption{}
\end{subfigure}
\end{center}
\caption{(Color online)  Time evolution of the expected number of excited qubits in the system on a generic 2 qubit system. The solid-blue line shows the dynamics when the system is subjected to incoherent pumping and the dashed yellow line when it is not. At $t=0$ the system starts with zero photons and the qubits in the state (a) $\ket{00}$; 
(b) $\ket{\Psi_+} = \dfrac{1}{\sqrt{2}}\left(\ket{01}+\ket{10}\right)$; and (c) $\ket{11}$.} \label{incexc}
\end{figure}

\medskip
\par
\begin{figure}[H]
\begin{center}

\includegraphics[width=0.50\textwidth]{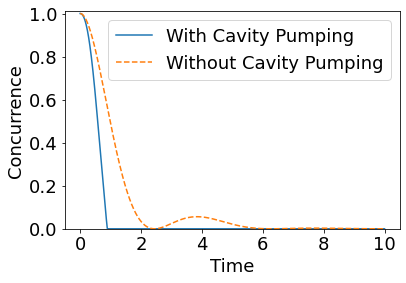}

\end{center}
\caption{(Color online)  Time evolution of the concurrence between the qubits, on a generic 2 qubit system. The solid-blue line shows the dynamics when the system is subjected to incoherent pumping and the dashed yellow line when it is not. At $t=0$ the system starts with zero photons and the qubits in the state  $\ket{\Psi_+} = \dfrac{1}{\sqrt{2}}\left(\ket{01}+\ket{10}\right)$.} \label{incconc}
\end{figure}

 
In order to illustrate the effects of incoherent pumping in the entanglement between qubits in the embedded in the cavity, we consider the simple case of 2 qubits with no qubit decay ($\gamma_q=0$). In this case, it was shown in Refs. that the maximally entangled state $\ket{\Psi_-} = \dfrac{1}{\sqrt{2}}\left(\ket{01}-\ket{10}\right)$ is protected from decay. We analyze the entanglement between the qubits by tracing out the cavity degrees of freedom and applying Eq. (\ref{conc}) to the reduced qubits density matrix at each time. The results are shown in Figs. \ref{incexc} \& \ref{incconc}, which compares  different aspects of the time evolution of the system  for the cases with incoherent cavity pumping and without it.  Each figure compares 4 different initial conditions: the system starts at $t=0$ always with zero photons, with the qubits starting in 3 different and orthogonal states ((a) $\ket{00}$; 
(b) $\ket{\Psi_+} = \dfrac{1}{\sqrt{2}}\left(\ket{01}+\ket{10}\right)$; and (c) $\ket{11}$). 
 All parameters were chosen arbitrarily.  Fig. \ref{incexc} shows the time evolution of the expected number of excited qubits, and Fig. \ref{incconc} shows the time evolution of the concurrence, the time evolution of the concurrence between the qubits, which was calculated by the application of Eq. (\ref{conc}).

\medskip
\par

Figure~ \ref{incexc} shows the time evolution of the number of excited qubits in a generic 2 qubit system for the system starting at $t=0$ in various states. Here we see that the addition of pumping in the system do create excitations in the qubits, as the expected value for the number of excited qubits when pumping takes place is always greater or equal to that when there is no pumping. This result is completely expected, since pumping adds photons to the cavity mode and photons in the cavity mode can, in turn, be consumed to create qubit excitations. We now turn our attention to the effects of incoherent pumping of photons to the actual quantum entanglement between the qubits.

\medskip
\par 
We readily see from Fig. \ref{incconc} that, even though incoherent pumping can create excitations in the qubits, it is unable to create entanglement between them. The quantum entanglement when incoherent pumping is present is always smaller or equal to that when it is absent, as witnessed by the concurrence between the qubits. We see that the state which was protected from entanglement decay in the case without pumping is still protected when there is pumping. Since our goal is to study entanglement creation and storage, from now on, we will focus entirely on coherent pumping of cavity photons. All other considered initial conditions showed no entanglement either created or destroyed by the dynamics. If the system starts the dynamics at either state $\ket{00}$ or $\ket{11}$, which are unentangled, the two qubits remain unentangled throuhout. If the system starts the dynamics on the state $\ket{\Psi_-} = \dfrac{1}{\sqrt{2}}\left(\ket{01}-\ket{10}\right)$, which is maximally entangled, it remains maximally entangled throughout.

\section{Coherent Pumping}\label{results}

 In this section, we analyze the effects of a coherent drive source of cavity photons in the entanglement dynamics between excitons in strained graphene. As discussed in Sec. \ref{realization}, such system can be described by the Tavis-Cummings Hamiltonian with the addition of a coupling term between the cavity and the pumping mode. For that, we numerically solved for the density matrix $\rho$ the master equation

\begin{equation}
\hbar \dot{\rho} = \dfrac{1}{i}\left[ \hat{H},\rho\right]  + \gamma_c\mathcal{L}(\hat{a}) \rho, \label{master3}
\end{equation}
where $\hat{H}$ is the modified Tavis-Cummings Hamiltonian, which is given in Eq. (\ref{hpump}). Throughout this section, we will calculate the concurrence, negativity, three-$\pi$, and mutual information using Eqs. (\ref{conc}), (\ref{negdef}), (\ref{threepidef}), and (\ref{MIdef}), respectively. We will also present an upper bound for the total negativity by the application of Eq. (\ref{Ntot}).

\medskip
\par 
We will now present and discuss the numerical results obtained through a simulation of the entanglement dynamics of excitons in strained graphene. We present our results in the following order: first, we do an in-depth study of the entanglement dynamics for a system of two excitons in strained graphene embedded in an optical microcavity and subject to a coherent pump under different conditions. In this section, we study how does the entanglement depends on the pumping rate, on the microcavity width, on the intensity of the strain-induced PMF, and on the product of the volume $W$ of the microcavity by its dielectric constant $\varepsilon_d$, which we relabel for simplicity as the factor $\alpha \equiv \varepsilon_d \times W$. We then consider systems with three excitons, in which we do a similar study of the dynamics of various entanglement witnesses for different values of the PMF, and of the constant $\alpha$. While studying systems of three excitons, we also compare the entanglement dynamics for different initial conditions of the system. In this  section we consider the following initial conditions for the system: the GHZ state, $\ket{GHZ} = \dfrac{1}{\sqrt{2}}\left(\ket{000}+\ket{111}\right)$ and zero cavity photons; the W state $\ket{W} = \dfrac{1}{\sqrt{3}}\left( \ket{001}+\ket{010}+\ket{100}\right)$ and zero cavity photons; and the N00N state $\ket{\Psi_{N00N}} = \dfrac{1}{\sqrt{2}}\left(\ket{0_c}\ket{111} +\ket{3_c}\ket{000}\right)$, where $\ket{n_c}$ is the cavity Fock state with $n$ cavity photons. Lastly, we study system with increasing number of excitons, and show the upper and lower bounds for the total negativity within the system as a function of the number of excitons for selected values of the PMF $B$ and of the factor $\alpha$. Throughout our simulations, unless specifically told otherwise, we considered the following physical parameters for the composed system: The intensity of the PMF in the graphene sheet was considered to be $B = 50$T; The microcavity's factor $\alpha = \varepsilon_d \times W$ was considered to be the same as the one of Ref. \cite{cav}, in which authors used an InGa microcavity ($\varepsilon_d \approx 13$) of volume $W = 1.69 \times 10^3 \mu$m$^3$ (we labeled the factor $\alpha$ for this microcavity to be $\alpha_0 \equiv 13 \times 1.69 \mu$m$^3$); The cavity width $L_C$ was considered to be such that the energy of the cavity photons is in resonance with the exciton energy gap ($\hbar\omega_C = \Delta_{ex}$); the cavity dedcay rate was considered to be $\gamma_C = 1.5$ GHz; and the coherent pumping rate was considered to be $R_P = 1.2$ GHz.


\subsection{Two Excitons}

 In this section, we will perform an in-depth study of the entanglement dynamics of a system of 2 excitons in strained graphene embedded in a leaky optical microcavity subject to a coherent pump of photons. We will study how this entanglement evolves on time and how does it depend on some physical parameters, such as the laser source pumping rate, the microcavity characteristic width, the intensity of the strain-induced PMF in the graphene sheet, and on the product of the volume $W$ of the microcavity  by it's relative dielectric constant $\varepsilon_d$. Here, we will only consider entanglement created when the system starts the dynamics in the ground state (both excitons in the ground state, and no cavity photons).

\medskip
\par
\begin{figure}[H]

\begin{center}
\begin{subfigure}[b]{0.32\textwidth}
\includegraphics[width=\textwidth]{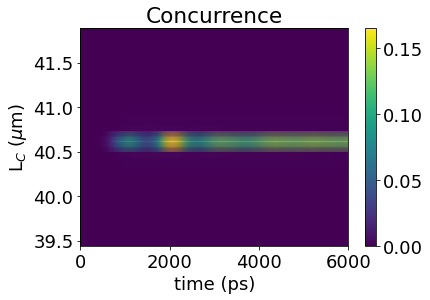}
\caption{}
\end{subfigure}
\begin{subfigure}[b]{0.32\textwidth}
\includegraphics[width=\textwidth]{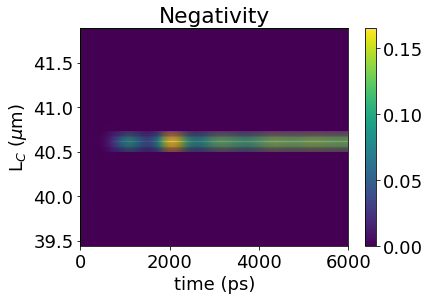}
\caption{}
\end{subfigure}
\begin{subfigure}[b]{0.32\textwidth}
\includegraphics[width=\textwidth]{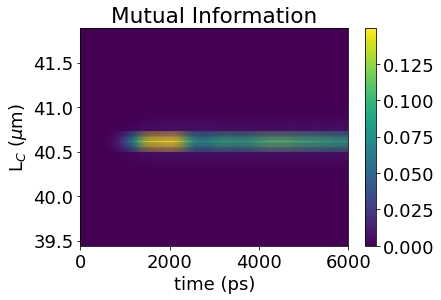}
\caption{}
\end{subfigure}
\end{center}
\caption{(Color online)  Quantum entanglement between two excitons in strained graphene as a function of the cavity width $L_C$, and of time. We considered the a InGa microcavity ($\varepsilon_d \approx 13$), the intensity of the strain-induced PMF to be $B = 50$ T, and the factor $\alpha$ to be $\alpha = \alpha_0$. Cavity photons are in resonance with the exciton binding energy when $L_C = 40.61$ $\mu$m. In (a) we see the Concurrence, in (b) the negativity, and in (c) the mutual information. \label{strainoff}}
\end{figure}

\begin{figure}[H]
\begin{center}
\begin{subfigure}[b]{0.49\textwidth}
\includegraphics[width=\textwidth]{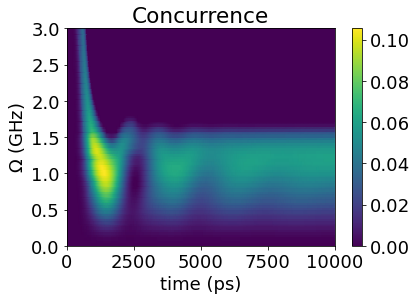}
\caption{}
\end{subfigure}
\begin{subfigure}[b]{0.49\textwidth}
\includegraphics[width=\textwidth]{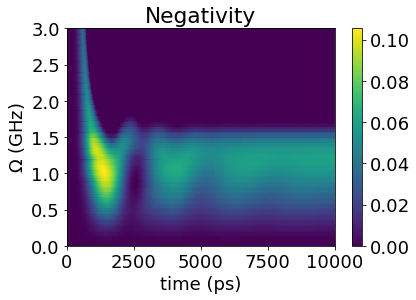}
\caption{}
\end{subfigure}\\
\begin{subfigure}[b]{0.49\textwidth}
\includegraphics[width=\textwidth]{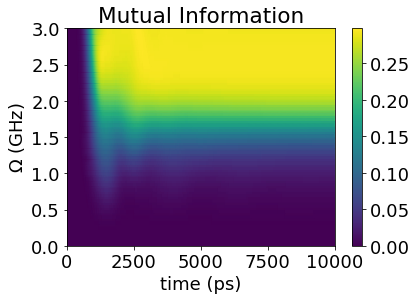}
\caption{}
\end{subfigure}
\begin{subfigure}[b]{0.49\textwidth}
\includegraphics[width=\textwidth]{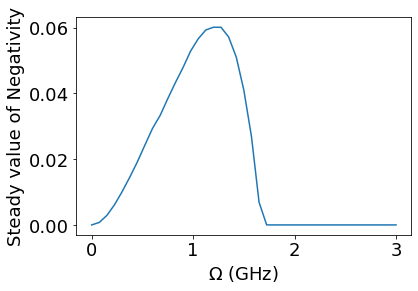}
\caption{}
\end{subfigure}
\end{center}
\caption{(Color online) Time evolution of the entanglement between two excitons in strained graphene as a function of the pumping rate $R_P$, and of time. In (a) we see the Concurrence, in (b) the negativity, and in (c) the mutual information. In (d) we see the value of the concurrence between the qubits when $t \rightarrow \infty$ as a function of the pump rate $R_P$. \label{coher}}
\end{figure}

\begin{figure}[H]
\begin{center}
\begin{subfigure}[b]{0.49\textwidth}
\includegraphics[width=\textwidth]{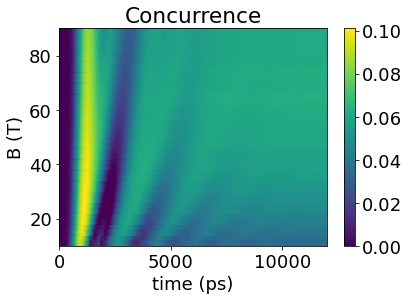}
\caption{}
\end{subfigure}
\begin{subfigure}[b]{0.49\textwidth}
\includegraphics[width=\textwidth]{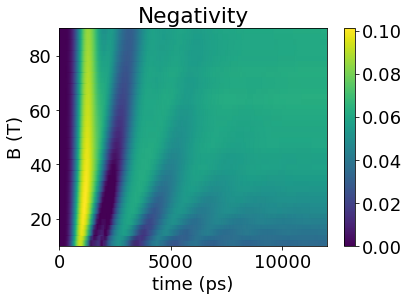}
\caption{}
\end{subfigure}\\
\begin{subfigure}[b]{0.49\textwidth}
\includegraphics[width=\textwidth]{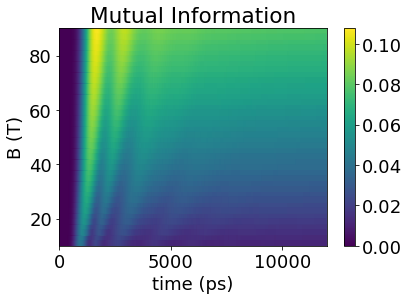}
\caption{}
\end{subfigure}
\begin{subfigure}[b]{0.49\textwidth}
\includegraphics[width=\textwidth]{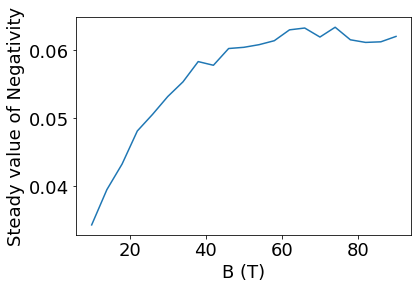}
\caption{}
\end{subfigure}
\end{center}
\begin{center}
\end{center}
\caption{(Color online)  Quantum entanglement between two excitons in strained graphene as a function of the intensity of the PMF $B$, and of time, when $\alpha = \alpha_0$. Here, at each trial, we consider that both the cavity mode and the laser are in resonance with the exciton binding energy. In (a) we see the Concurrence, in (b) the negativity, and in (c) the mutual information. In (d), we see the steady value of the negativity as a function of the PMF $B$. \label{strainres}}
\end{figure}

\begin{figure}[H]
\begin{center}
\begin{subfigure}[b]{0.49\textwidth}
\includegraphics[width=\textwidth]{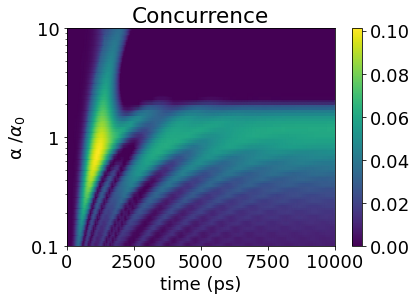}
\caption{}
\end{subfigure}
\begin{subfigure}[b]{0.49\textwidth}
\includegraphics[width=\textwidth]{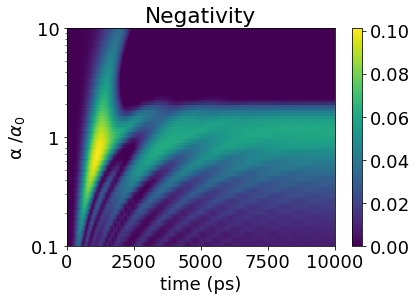}
\caption{}
\end{subfigure}\\
\begin{subfigure}[b]{0.49\textwidth}
\includegraphics[width=\textwidth]{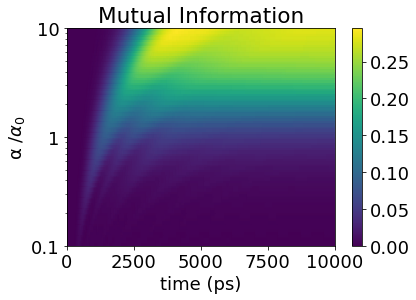}
\caption{}
\end{subfigure}
\begin{subfigure}[b]{0.49\textwidth}
\includegraphics[width=\textwidth]{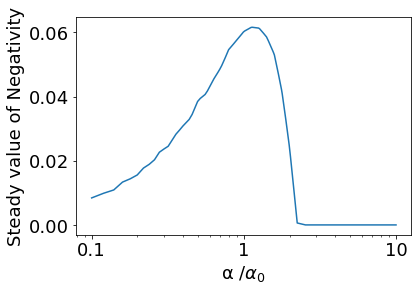}
\caption{}
\end{subfigure}
\end{center}
\begin{center}
\end{center}
\caption{(Color online) Quantum entanglement between two excitons in strained graphene as a function of the microcavity's parameter $\alpha\equiv \varepsilon_d\times W$
, and of time, when $B = 50$ T. Here, at each trial, we consider that both the cavity mode and the laser are in resonance with the exciton energy gap. In (a) we see the Concurrence, in (b) the negativity, and in (c) the mutual information. In (d), we see the steady value of the negativity as a function of $\alpha$. \label{alpha2}}
\end{figure}

 
The results in this section are presented as following: In Fig. \ref{strainoff}, we study the dependence the entanglement dynamics on the cavity width $L_C$, and show that lasting entanglement can only be observed when the cavity photons are in resonance with the excitonic energy gap. Fig \ref{coher} shows how does the entanglement dynamics varies when we change the rate $R_P$ at which the microcavity is being pumped with photos. In Fig. \ref{strainres}, we study how the entanglement dynamics changes when we change the intensity of the pseudomagnetic field $B$. Similarly, Fig. \ref{alpha2} depicts the influence of the value of the microcavity's  $\alpha$ factor on the entanglement dynamics. 

\medskip
\par 
In Fig. \ref{strainoff}, we see the effect of changing the microcavity width $L_C$ in the resulting entanglement when  the intensity of the strain-induced PMF is equal to $B/e = 50$ T. We see that, in this case, entanglement is only created very close to the resonant value of $L_C$, which is $L_{res} = 40.61$ $\mu$m. Entanglement decays to zero very rapidly when the width is even slightly off-resonance. This means that, in order to effectively monitor entanglement, the microcavity has to be very precisely engineered to be in resonance with the qubits when they are under the desired strain and cannot be used for even slightly different strain intensities.


\medskip
\par 

We see from Fig. \ref{coher} that providing an external coherent source of photons to the cavity does create some entanglement between the qubits. We see that this entanglement, however, is not long-lasting if the pump source is too strong. We see that, if the pumping rate $R_P$ is greater than the cavity decay rate $\gamma_c$, the entanglement will rapidly decay to zero after a small peak at the beginning of the dynamics. This can be understood by the fact that, in this regime, photons are being pumped in the cavity more rapidly than they can decay which leads to a big number of cavity states being accessible. When the cavity degrees of freedom are traced out to calculate the concurrence between the qubits, the many non-zero Fock states amount to a significant loss of information, whereas when only a few states are accessible by the photons, the loss of information from tracing the cavity out is significantly smaller.

\medskip
\par 
In Fig. \ref{strainres}, we study how the entanglement created between the excitons depend on the PMF $B$. We see that the interaction with the cavity photons creates some early peaks of entanglement that oscillates in time until it reaches a final steady value. In Fig. \ref{strainres} (d), we see that this final steady value of entanglement increases with the intensity of the PMF $B$. 


\medskip
\par 
The results shown in Fig. \ref{alpha2}, show us how does the cavity volume and dielectric constant effect the entanglement created between the excitons. In this figure, we consider changing the parameter $\alpha$ of the microcavity while maintaining the microcavity resonance frequency $\omega_C$ unchanged. We know from Eq.(\ref{rabistrain}) that the Rabi coupling strength $g$ between excitons and cavity photons is inversely proportional to the square root of alpha, $g\propto \dfrac{1}{\sqrt{\alpha}}$. Effectively, this changes the intensity $g$, while maintaining the other parameters of the Master Equation (\ref{master_pump}) unchanged.  In Fig. \ref{alpha2} (d), we see a peak in the entanglement between the two excitons appears at around $\alpha \approx \alpha_0$, which sharply decays to zero entanglement after a cutoff threshold that, for our chosen parameters, happens around $\alpha \approx 2\alpha_0$. This means that there exists a minimum value for the Rabi coupling $g = g_{\mathrm{min}}$, which, if $g<g_{\mathrm{min}}$, the excitons do not get entangled with each other at long times (although some entanglement does appear before the steady-state is reached, as can be seen in Figs. \ref{alpha2} (a) and (b)). An interesting result is that arbitrarily increasing the intensity of the coupling between excitons and cavity photons, however, ends up decreasing the lasting entanglement between the excitons themselves. This happens when the value of $\alpha$ decreases, as can be seen in Fig. \ref{alpha2} (d). 

\subsection{Three Excitons}

We now consider a system of three excitons in strained graphene. We divide our results into two parts. In the first part, we calculated the time evolution of the negativity between one exciton and the remaining two excitons, the three-$\pi$ of the combined system, the concurrence between a pair of excitons (when one of the excitons is traced out), and the steady value of the negativity.  In Fig. \ref{3qbts}, we study how those traditional measures of entanglement depend on the intensity of the PMF $B$, and on Fig. \ref{3qbtsalpha}, we study how do they depend on the cavity's volume and dielectric constant through the parameter $\alpha$. In the second part, we study different initial conditions for the system. Here, we compare the dynamics of the negativity between one exciton and the remaining two excitons for four different initial conditions. 

\subsubsection{Entanglement created from the ground state}

\begin{figure}[H]
\begin{center}
\begin{subfigure}[b]{0.40\textwidth}
\includegraphics[width=\textwidth]{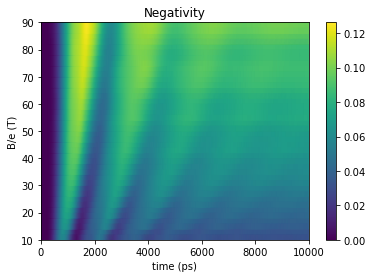}
\caption{}
\end{subfigure}
\begin{subfigure}[b]{0.40\textwidth}
\includegraphics[width=\textwidth]{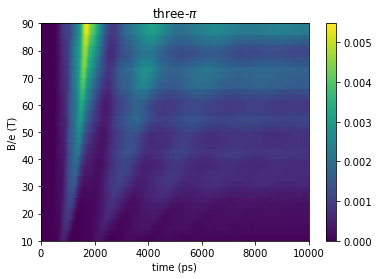}
\caption{}
\end{subfigure}\\
\begin{subfigure}[b]{0.40\textwidth}
\includegraphics[width=\textwidth]{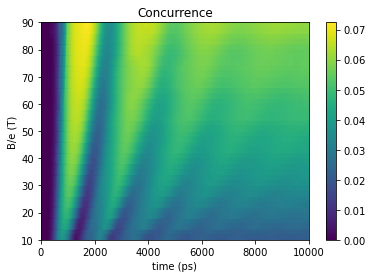}
\caption{}
\end{subfigure}
\begin{subfigure}[b]{0.40\textwidth}
\includegraphics[width=\textwidth]{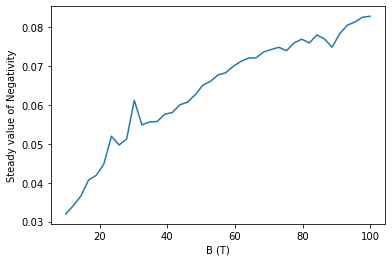}
\caption{}
\end{subfigure}
\end{center}
\begin{center}
\end{center}
\caption{(Color online) Quantum entanglement in a three-exciton system in strained graphene as a function of the intensity of the PMF $B$ and of time, when $\alpha = \alpha_0$. In (a) we see the negativity, in (b) the three-$\pi$, in (c) we see the concurrence between two of the excitons, and in (d) we see the steady value of the negativity at long times as a function of the PMF $B$. \label{3qbts}}
\end{figure}

\medskip
\par 
In Fig. \ref{3qbts}, we show how does the intensity of the PMF $B$ influence the dynamics of different measures of entanglement. From it's panels, we reach similar conclusions as those obtained from those of Fig. \ref{strainres}. We see that, long-lasting entanglement is still created. We also see that this entanglement increases with the intensity of the PMF $B$. One important result from Fig. \ref{3qbts} can be seen in panel (b). In Fig. \ref{3qbts} (b), we see that the interaction of the system with the laser creates some entanglement that is shared by all three excitons, as can be seen by a final non-zero value of the three-$\pi$. This entanglement, however, is extremely weak when compared to the overall value of the negativity. This implies that the majority of the entanglement created in this composed system is shared only by subsets of two excitons. Other interesting results comes from the comparison of panels (c) and (d) with panels (a) and (d) of Fig. \ref{strainres}. By doing so, we note that the overall amount of entanglement each exciton experiences has increased, as represented by an increase in the value of the steady value of the negativity, but the entanglement experienced by a given pair of excitons decreases, as the concurrence between two excitons is smaller in the three-exciton system than in the two-exciton system. That being said, as we will show when we study larger systems, it is not always true that increasing the number of excitons increases the overall amount of entanglement.  


\begin{figure}[H]
\begin{center}
\begin{subfigure}[b]{0.40\textwidth}
\includegraphics[width=\textwidth]{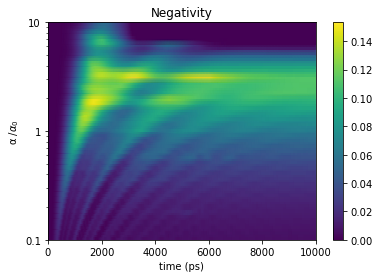}
\caption{}
\end{subfigure}
\begin{subfigure}[b]{0.40\textwidth}
\includegraphics[width=\textwidth]{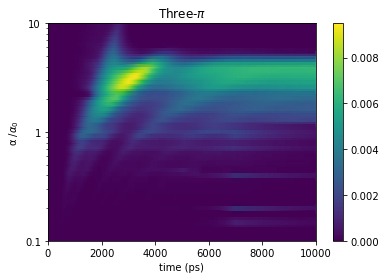}
\caption{}
\end{subfigure}\\
\begin{subfigure}[b]{0.40\textwidth}
\includegraphics[width=\textwidth]{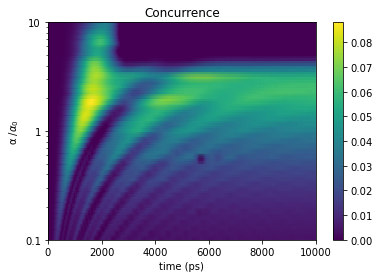}
\caption{}
\end{subfigure}
\begin{subfigure}[b]{0.40\textwidth}
\includegraphics[width=\textwidth]{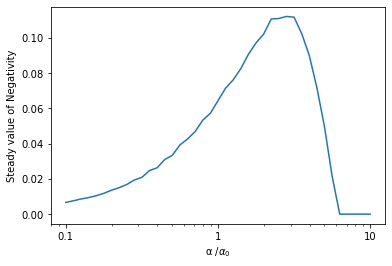}
\caption{}
\end{subfigure}
\end{center}
\begin{center}
\end{center}
\caption{(Color online) Quantum entanglement in a three-exciton system in strained graphene as a function of the parameter $\alpha$ and of time, when $B$= 50 T. In (a) we see the negativity, in (b) the three-$\pi$, in (c) we see the concurrence between two of the excitons, and in (d) we see the steady value of the negativity at long times as a function of $\alpha$. \label{3qbtsalpha}}
\end{figure}

\medskip
\par 
At a first glance, the results shown in Fig. \ref{3qbtsalpha}, might appear to be completely akin to those of Fig. \ref{alpha2}. We see the same overall pattern of a peak in the entanglement when $\alpha$ reaches an optimal value $\alpha = \alpha_{\mathrm{opt.}}$, with the entanglement decreasing as $\alpha \rightarrow 0$, and  with a sharp transition to zero entanglement at some cutoff threshold. Like for the two exciton case, we can understand this when we remember that the Rabi coupling decreases with the square root of $\alpha$. If the Rabi coupling is too small (if $\alpha$ is larger than the cutoff threshold), no entanglement is created in the composed system. What is very interesting, however, is that the values of the cutoff threshold and of the optimal alpha for maximum entanglement are significantly different for both cases. When we look at Fig. \ref{alpha2} (d), we see that the optimal value for alpha on that case is $\alpha_{\mathrm{opt.}}\approx \alpha_0$ and the cutoff tharshold happens at around $\alpha_\mathrm{max} = 2.0 \alpha_0$. If we look at Fig. \ref{3qbtsalpha} (d), however, we see that, for three excitons, these values are around $\alpha_{\mathrm{opt.}}\approx 2.5 \alpha_0$, and $\alpha_\mathrm{max} = 7.0 \alpha_0$. This means that the peak of entanglement observed by a system of 3 excitons happens at a point in which a system of only 2 excitons exhibit no entanglement whatsoever. 

\subsubsection{Entanglement created from different initial conditions}

So far, we have only considered how coherent pumping a cavity can create entanglement in a system initially in the ground state. That is, however, not the only relevant situation. One might want to know how initially entangled states evolve under such circumstances. We will now compare four different initial conditions in a system of 3 qubits. In order to conduct such analysis, we chose some of the most usual states when studying the entanglement between three qubits. We will compare how quantum entanglement between the qubits evolves when the system starts in the ground state; when the cavity starts with zero photons and the qubits start in both the $W$ state $\ket{W} = \dfrac{1}{\sqrt{3}}\left(\ket{001}+\ket{010}+\ket{100}+\right)$, and the $GHZ$ state $\ket{GHZ} = \dfrac{1}{\sqrt{2}}\left(\ket{000}+\ket{111}\right)$; and when the combined system starts in the entangled N00N state, which is a linear superposition of the state in which the cavity has 3 photons and no excited qubits, and the state in which the cavity has zero photons, and all qubits are excited, namely $\ket{n00N} = \dfrac{1}{\sqrt{2}}\left(\ket{0_ph}\ket{111} + \ket{3_ph}\ket{000}\right)$, where $\ket{n_ph}$ is the cavity Fock state with $n$ photons. 

\begin{figure}[H]
\begin{center}
\begin{subfigure}[b]{0.30\textwidth}
\includegraphics[width=\textwidth]{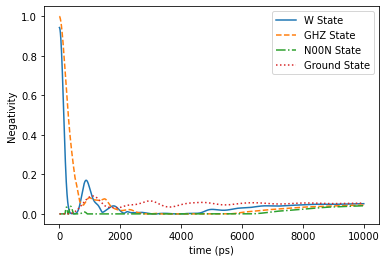}
\caption{}
\end{subfigure}
\begin{subfigure}[b]{0.30\textwidth}
\includegraphics[width=\textwidth]{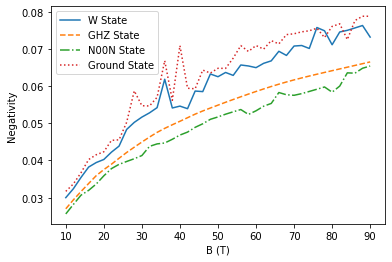}
\caption{}
\end{subfigure}
\begin{subfigure}[b]{0.30\textwidth}
\includegraphics[width=\textwidth]{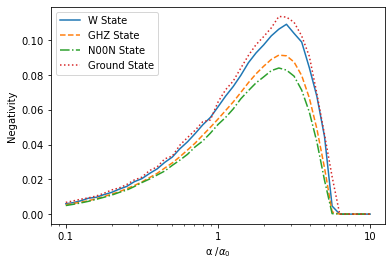}
\caption{}
\end{subfigure}
\end{center}
\caption{(Color online) Comparison of the time evolution of the Negativity for four different Initial conditions. The solid blue line represents the system starting in the $W$ state; the dashed yellow line represents the system starting in the $GHZ$ state; the dot-dashed green line represents the system starting in the $N00N$ state; and the dotted red line the system starting in the ground state. On (a) we see the Negativity between the qubits as a function of time for the case in which $B = 50$ T, and $\alpha = \alpha_0$. In (b) we see the steady value of the negativity as a function of the PMF $B$, when $\alpha = \alpha_0$. In (c), we see the steady value of the negativity as a function of the parameter $\alpha$, when $B$ = 50 T.\label{states}}
\end{figure}

Each of these states is interesting for particular reasons: The $GHZ$ state is an entangled state in which entanglement can only be seen when analyzing all of the entangled qubits. If someone was to study only one pair of qubits, the resulting density matrix for the qubits 1 and 2 would be $\rho_{1,2} = \texttt{Tr}_3 \ket{GHZ}\bra{GHZ} = \dfrac{1}{2}\left(\ket{00}\bra{00} + \ket{11}\bra{11}\right)$, which is a simple mixture of separable states and shows, therefore, no entanglement. The $W$ state, on the other way, has most of it's entanglement shown between pairs of qubits. This can be seen by the fact that the three-tangle, which is used to measure only three-way entanglement, is zero for this state. However, one cannot say that the $W$ state has absolutely no entanglement shared three-ways, since it has non-zero three-pi, another measure of three-way entanglement. The N00N state is a state in which the qubits are entangled to the cavity itself. Studying the qubits alone in such a system yields no entanglement, as the reduced matrix in the qubits subspace is simply $\rho_{qb} = \dfrac{1}{2}\left(\ket{000}\bra{000}+\ket{111}\bra{111}\right)$. Our results are shown in Fig. \ref{states}.

Looking at Fig. \ref{states}, we see that, overall, all considered initial states evolve to a similar final state with a small, but non-zero, value for the lasting entanglement.  The steady value of the negativity also shows only small variations for different initial conditions. The overall pattern of how this steady value of negativity depends on both the pseudomagnetic field $B$ - seen on Fig. \ref{states} (b) - and on the factor $\alpha$ - seen on Fig. \ref{states} (c) - are virtually the same for all considered initial conditions.

\subsection{Multi-Excitonic systems}

As we can see from Figs. \ref{strainres} and \ref{3qbts}, providing a coherent source of photons to the microcavity creates lasting entanglement between the excitons, even in imperfect cavities.  We will now study how this entanglement changes when more excitons are added to the system.



Since the dynamics governed by Eq. (\ref{master_pump}), and all considered initial states are symmetric under an exchange of qubits, the negativities with respect to each individual qubits are all the same and equal to $\mathcal{N}_j \equiv \mathcal{N}$, $\forall j\in \{1,\dots,N\}$. In this case, the upper and lower bounds for the total negativity, given by Eq. (\ref{Ntot}), become
\begin{equation}
\mathcal{N}\leq \mathcal{N}_{tot} \leq \dfrac{N}{2} \mathcal{N}. \label{Ntot2}
\end{equation}

\begin{figure}[H]
\begin{center}
\begin{subfigure}[b]{0.30\textwidth}
\includegraphics[width=\textwidth]{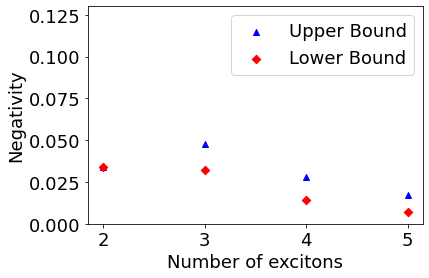}
\caption{B = 10 T}
\end{subfigure}
\begin{subfigure}[b]{0.30\textwidth}
\includegraphics[width=\textwidth]{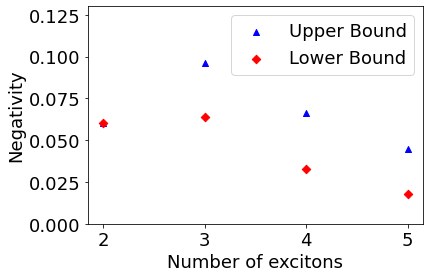}
\caption{B = 50 T}
\end{subfigure}
\begin{subfigure}[b]{0.30\textwidth}
\includegraphics[width=\textwidth]{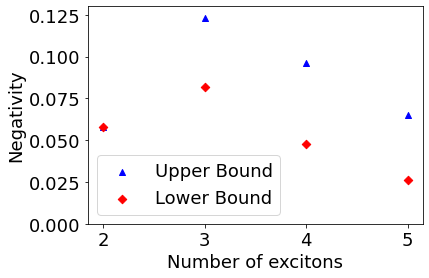}
\caption{B = 100 T}
\end{subfigure}
\end{center}
\caption{(Color online) Upper (blue triangle) and lower (red diamond) bounds for the total negativity in a 10 ns simulation as a function of the number of excitons coupled to the optical microcavity, up to a total of 6 excitons, in a strained sheet of graphene for different values of the PMF $B$, in the case in which $\alpha = \alpha_0$. \label{negublb}}
\end{figure}

\begin{figure}[H]
\begin{center}
\begin{subfigure}[b]{0.30\textwidth}
\includegraphics[width=\textwidth]{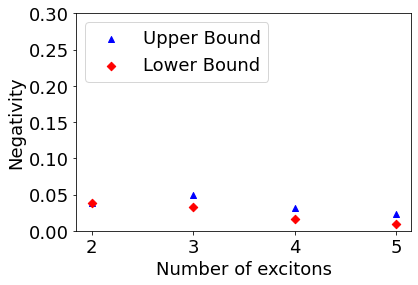}
\caption{$\frac{\alpha}{\alpha_0} = 0.5$}
\end{subfigure}
\begin{subfigure}[b]{0.30\textwidth}
\includegraphics[width=\textwidth]{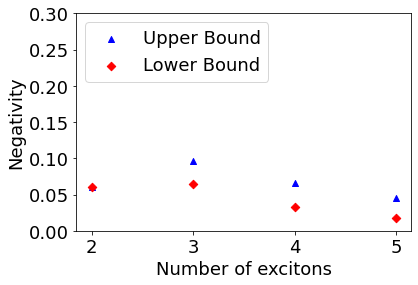}
\caption{$\frac{\alpha}{\alpha_0} = 1.0$}
\end{subfigure}\\
\begin{subfigure}[b]{0.30\textwidth}
\includegraphics[width=\textwidth]{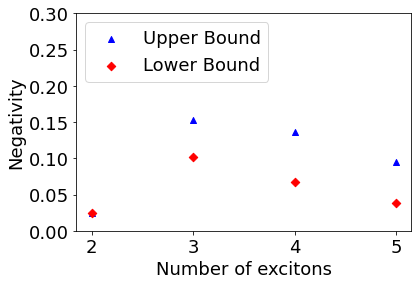}
\caption{$\frac{\alpha}{\alpha_0} = 2.0$}
\end{subfigure}
\begin{subfigure}[b]{0.30\textwidth}
\includegraphics[width=\textwidth]{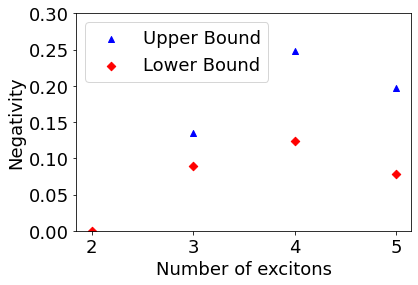}
\caption{$\frac{\alpha}{\alpha_0} = 4.0$}
\end{subfigure}
\end{center}
\caption{(Color online) Upper (blue triangle) and lower (red diamond) bounds for the total negativity in a 10 ns simulation as a function of the number of excitons coupled to the optical microcavity, up to a total of 6 excitons, in a strained sheet of graphene for different values of the parameter $\alpha$, in the case in which $B$ = 50 T.  \label{negublbalpha}}
\end{figure}

 We will now show the upper and lower bounds of Eq. (\ref{Ntot2}) for different values of the PMF - in Fig.   \ref{negublb} - and of the parameter $\alpha$, in Fig. \ref{negublbalpha}.

As it can be seen from Fig.  \ref{negublb}, a very interesting and counterintuitive result was reached.  Freely adding additional excitons to the system appears to have the effect of diminishing the overall amount of entanglement contained within it. The entanglement reaches a maximal value when 3 excitons are present, before a rapid decay as more excitons are added.

\medskip
\par 
In Fig. \ref{negublbalpha}, we see the continuation of the trend we observed with Figs. \ref{alpha2} and \ref{3qbtsalpha}. We see that increasing the value of $\alpha$ can increase the overall amount of entanglement for multi-excitonic systems. In the plots shown, for example, the maximum amount of entanglement happens for a system of 4 excitons in strained graphene when $\alpha = 4 \alpha_0$. At this point, we can see that a system of only 2 excitons shows no entanglemetn whatsoever, and a system of 3 excitons shows less entanglement than it did when $\alpha = 2\alpha_0$, for example. This means that, in order to maximize the amount of entanglement in such systems, the parameters of the cavity must be adjusted for the exact number of excitons within the graphene sheet.

\section{Model Limitations and interpretation of results \label{limitations}}

Let us now consider the interaction between excitons. It is non-monotonic and similar in form to the Lenard-Jones potential, which describes, for example, the interaction of two hydrogen atoms - an analogue of excitons. At large separations $r$, the interaction between pseudo-magnetoexcitons is a van der Waals attraction, and at small separations on the order of the exciton radius $R$, the interaction becomes repulsive due to the overlap of excitons. Therefore, with increasing density $n$, there is an increase in attraction, which at densities of the order of $\sim 1/\pi R^2$ is replaced by repulsion. The Tavis-Cummings model employed in this work takes into consideration the interactions between excitons and one optical mode in the optical cavity.

\medskip
\par 
Next, let us consider the scope of applicability of this model. This model does not take into account interactions between qubits. It does not take into account the specifics of entanglement of distant qubits \cite{lim1} and does not take into account the imperfection of the optical cavity \cite{lim2}. In addition, the Tavis-Cummings Model in our work is not used to describe the dynamics of qubit entanglement at short times, that is, it does not describe non-Markov oscillations. It describes the dynamics of qubit entanglement only at large times, in the Markov region.
Due to the preceding remarks, the certainly important and useful results obtained within the Tavis-Cummings Model are applicable to describe the dynamics of entanglement in the region of control parameters (concentration, times, etc.) where they do not contradict the physical implementation used.

\medskip
\par 
Let us consider, for example, the dependence of the entanglement on the pseudomagnetic field $B$. When the field tends to zero, the distance between Landau levels tends to zero and excitons can no longer be considered as qubits. This occurs when the distance between Landau levels (it depends linearly on $B$ for gapped graphene and as the square root of $B$ for gapless graphene) becomes smaller than all other characteristic energies - such as the thermal energy $kT$ ($T$ is the absolute temperature). Therefore, for such magnetoc fields $B$, the Tavis-Cummings model used in the calculations is not applicable.

\medskip
\par
When exciton-exciton interactions are considered, one should expect that, as the concentration is decreased, the probability of entanglement between distant qubits is decreased, and with increasing density it increases. However, when the concentration of qubits is high and each qubit is surrounded by neighbors, it can only become entangled with them, but not with more distant ones. Consequently, at concentrations of the order of $\sim 1/\pi R^2$, where $R$ is the radius of excitons, entanglement should begin to decrease.

\medskip
\par 
Interestingly enough, our results show the same pattern, even though exciton-exciton interactions are neglected in our calculations. We see a maximum in the quantum entanglement at a finite value of the factor $\alpha$, which is proportional to the cavity volume. The effect due to increasing the factor $\alpha$ is to, effectively decreasing the excitonic density $n$ within the microcavity, since $n = N/W$, and $\alpha \propto W$. If we expect maximum entanglement to occur at an optimal value for the excitonic density, we should see systems with a larger number of excitons maximizing their entanglement at higher values of $\alpha$ than  systems with smaller number of excitons. This is exactly what was observed in our simulations.

\section{Discussion and Conclusions}\label{conclusion}

In this paper, we have studied the dynamics of quantum entanglement of multi-qubit systems embedded in leaky optical microcavities. We  considered a system of excitons in strained graphene whose energy gap between the ground and first excited state is either in, or near resonance to a cavity mode, as well as a generic system of 2 qubits being subjected to an incoherent source of cavity photons.

\medskip
\par 
Our proposed system enables achieving quantum entanglement of excitons at room temperatures due to large exciton binding energy in 2D materials.  The importance of quantum entanglement of three and more qubits for quantum computing is related to the fact that the entangled state of three qubits is characterized by higher fidelity than the entangled state of two qubits \cite{fidelity}. In addition, multi-qubit entanglement is very useful for quantum error correction codes \cite{Shor,Steane}.

\medskip
\par 
The first result we obtained was not very surprising. We showed that for a generic system of qubits coupled to a leaky cavity which is being pumped by an incoherent source of photons, even though excitations can and do appear in the qubits, entanglement between them is never established. The incoherent source of photons actually does the opposite. It speeds up the rate in which entanglement is destroyed in most cases (the only exception being one maximally entangled state that was protected from decay in the case without pumping, which was still protected from decay, even when pumping is present). This is, again, not very surprising, since decoherence is a well known mechanism of entanglement decay \cite{ann}.

\medskip
\par 
After this brief analysis, we studied the effects due to a coherent source of cavity photons in the semiclassical limit, in which the outside pumping source is treated classically while the cavity mode is still treated by quantum electrodynamics. We then turned our attention to a system of excitons in strained graphene, embedded in an optical microcavity. We first studied the effects of increasing the frequency at which photons are being pumped in the system. There we reached the counter-intuitive conclusion that, although a coherent source of photons do create quantum entanglement between the excitons, increasing the frequency at which photons are being pumped to the system eventually destroys any long-living entanglement between them. This happens whenever the rate at which photons are pumped becomes greater than the decay rate of the cavity. In this regime, all cavity Fock states become effectively viable, and are eventually populated as time runs.  Therefore, in this regime, the microcavity gains access to a large number of degrees of freedom (one for each accessible Fock state). When we trace out the cavity degrees of freedom in order to evaluate the entanglement between the excitons themselves, the loss of information becomes too large and no entanglement can be observed between the excitons alone. 
On the other hand, when the decay rate is greater than the pumping rate, only a few Fock st tes are effectively populated, and the overall amount of information lost when the cavity is traced out is much smaller, resulting in long-living entanglement between the excitons. 

\medskip
\par 
We studied the effect of the microcavity width in the overall entanglement between 2 excitons. We have shown that entanglement is only formed when the width of the cavity is such as the photons confined to it are in resonance with the energy gap in the excitons. Even small deviations of half a percent in the energy of the photons from the exciton energy gap are sufficient to destroy the resulting entanglement almost entirely. This means that any experimental realization would need to be very precise in applying the exact amount of pressure in the graphene sheet as to make excitons in it to be in resonance with the energy of cavity photons.

\medskip
\par 
Before considering systems with more than two qubits, we studied the effect of increasing the applied pseudomagnetic field in the graphene sheet on the overall entanglement between the excitons. There, our results showed that for more intense fields, we obtain more entangled excitons. We also showed that the excitons will reach their final entanglement faster in less intense fields.  We have also studied how the entanglement depends on the overall volume and dielectric constant of the microcavity. Changing those parameters while maintaining the frequency of cavity photons unchanged, effectively changes the Rabi coupling between cavity photons and excitons, while maintaining all other parameters of the master equation that governs the dynamics fixed. We saw that, for cavities with larger volume, in which the Rabi coupling is weaker, the entanglement decays sharply to zero after a cutoff threshold. This means that, if the Rabi coupling is weaker than a minimum value, no lasting entanglement is created between the excitons. Interestingly, the amount of entanglement is not a monotonic function of the Rabi coupling. The system shows maximum entanglement at a finite value of the Rabi coupling. This can be understood by the fact that large coupling leads to very fast interactions between excitons and photons, which can average out to zero in the extreme limit, drastically reducing the effects of those interactions in a similar fashion to the rotating wave approximation.

\medskip
\par 
For systems with three excitons, we conducted the same analysis on the dependence of the entanglement created from the ground state on the intensity of the pseudomagnetic field and reached the same conclusion: with more intense fields we have more entanglement, and slower reaching of the steady value for the entanglement measures. During this study, we calculated the value of the three-$\pi$ of the system and showed that, although there is some entanglement shared three-ways, as evidenced by the non-zero three-$\pi$, most of the entanglement in the system is shared by pairs of excitons, as the overall negativity was significantly greater than the three-$\pi$. When studying the dependency of the entanglement on the volume and dielectric constant of the microcavity, we reached the same conclusion as for the case of two excitons, with one important difference: both the cutoff threshold, after which no lasting entanglement can be seen and the optimal value for the factor $\alpha$, defined as the product of the microcavity volume by it's dielectric constant, were significantly higher for three excitons than for two excitons. Interestingly enough, our results showed that, in order for a system of three excitons to experience maximum entanglement, the optimal value of the parameter $\alpha$ is actually above the cutoff threshold for two-exciton systems. This means that a system of three excitons would show maximum entanglement in the same settings that would prevent a system of two excitons of showing any entanglement whatsoever.

\medskip
\par 
We also conducted a study on the entanglement dynamics for systems that are originally entangled. We compared the dynamics with the excitons starting in the GHZ state, $\ket{GHZ} = \dfrac{1}{\sqrt{2}}\left(\ket{000}+\ket{111}\right)$, in the W state, $\ket{W} = \dfrac{1}{\sqrt{3}}\left(\ket{001}+\ket{010}+\ket{100}\right)$, in the ground state, all of which the system started with zero cavity photons, and  with the whole system starting in the N00N state $\ket{n00N} = \dfrac{1}{\sqrt{2}}\left(\ket{0_ph}\ket{111} + \ket{3_ph}\ket{000}\right)$. Our results show that all considered initial conditions evolved to final states with similar value for the negativity and that this final value for the negativity depends on $B$ and on $\alpha$ in a very similar fashion.  The results obtained here are undoubtedly applicable to a wider range of physical implementations than the physical system we focus on in our work— excitons in a strained graphene monolayer. Similar conclusions should be reached for any system that can be adequately approximated by the Tavis-Cummings model.

\medskip
\par 
Lastly, we analyzed how the final value of the negativity between one exciton and the remaining excitons changes with the total number of excitons. We've shown this value for up to five total excitons.  We saw the counter-intuitive result that adding more excitons to the system does not necessarily increases the amount of entanglement created in the system. We saw that, for each value of the $\alpha$ parameter, there is an optimal amount of excitons, in which the overall amount of entanglement is maximal. After we reach this optimal amount, adding more excitons to the system ends up reducing the overall amount of entanglement shared by the excitons.

\medskip
\par 
We propose a new possible realization for quantum technologies based on excitons in strained 2D materials. The advantage of use of these excitons for quantum technologies is possible operation at temperatures significantly higher than the temperatures at which quantum computers operate on superconducting qubits (see Refs.~\cite{Fogler1,Fogler2}). Harnessing  excitons in two-dimensional materials~\cite{Fogler1,Fogler2} for qubits mak s quantum computing to be possible even at room temperature. This is due to the fact that excitons are not destroyed at temperatures below their binding energies. The temperature of condensation of two-dimensional excitons is directly proportional to their density. The maximal available density is inversely proportional to square of the exciton radius, which is very small for new 2D materials. Therefore, the condensation of excitons in 2D materials is
achievable at room temperatures.

\medskip
\par

Proposed studies of multi-qubit entanglement between excitons in  strained 2D materials at room temperatures promise to be very important for the development of quantum communication, quantum cryptography and quantum computation.  Our results will showcase the potential of 2D material-based exciton qubit platforms for multiqubit quantum algorithms.


\section*{Acknowledgments}

The authors are grateful to M. Amico and O. V. Roslyak for valuable discussions. The authors are grateful for support by grants:  O.L.B.  acknowledges the support from the PSC-CUNY Award $\#$ 66382-00 54. Yu.E.L. acknowledges the support from the RSF  23-12-00115.


\begin{thebibliography}{99}

\bibitem{Nielsen_book} M. A. Nielsen and I. L. Chuang, \textit{Quantum Computation and Quantum Information}, (Cambridge Univ. Press, 2000). 

\bibitem{Horodecki} R. Horodecki, P. Horodecki, M. Horodecki, and K. Horodecki, Quantum Entanglement \rmp \textbf{81}, 865 (2009).

\bibitem{conc} W. K. Wooters, Entanglement of Formation of an Arbitrary State of Two Qubits, \prl \textbf{80}, 2245 (1998)

\bibitem{tangle} V. Coffman, J. Kundu, and W. K. Wootters, Distributed Entanglement, \pra \textbf{61}, 052306 (2000).

\bibitem{mconc} F. Mintert, M. Kuś, and A. Buchleitner, Concurrence of Mixed Multipartite Quantum States, \prl \textbf{95}, 260502 (2005).

\bibitem{nega} G. Vidal and R. F. Werner, Computable Measurements of Entanglement, \pra \textbf{65}, 032314 (2002).

\bibitem{threepi}  Y. C. Ou, and H. Fan, Monogamy inequality in terms of negativity for three-qubit states, \pra \textbf{75}, 062308 (2007).

\bibitem{mutual}  M. B. Plenio and S. Virmani, An introduction to entanglement measures. \textit{Quantum Inf. Comput.} \textbf{7}, 1 (2007).

\bibitem{Penrose} R. Penrose,  Quantum computation, entanglement and state reduction. \textit{Phil. Trans. R. Soc. A} \textbf{356} 1927–1939 (1998).

\bibitem{Jozsa} R. Jozsa, and N. Linden, On the role of entanglement in quantum-computational speed-up. \textit{Proc. R. Soc. Lond. A} \textbf{459} 2011–2032 (2003).

\bibitem{Zou} N. Zou, Quantum Entanglement and Its Application in Quantum Communication, J. Phys.: Conf. Ser. \textbf{1827} 012120 (2021).



\bibitem{EPR} A. Einstein, B. Podolsky, and N. Rosen, Can quantum-mechanical description of physical reality be considered complete? \textit{Physical Review}, \textbf{47} (10), 777 (1935).

\bibitem{Bell} J.S. Bell, On the Einstein Podolsky Rosen paradox. \textit{Physics Physique Fizika} \textbf{1} (1964).


\bibitem{ann} K. Ann, and G. Jaeger, Finite-time destruction of entanglement and non-locality by environmental influences, \textit{Foundations of Physics} \textbf{39}  790-828 (2009).

\bibitem{yu} T. Yu, and  J.H. Eberly, Sudden death of entanglement: Classical noise effects. \textit{Optics Communications} \textbf{264}, 2, pp 393-397 (2006).

\bibitem{Almeida} M. P. Almeida \textit{et al.} Environment-Induced Sudden Death of Entanglement. \textit{Science} \textbf{316}, 579-582 (2007).

\bibitem{Eberly} J. H. Eberly, and Ting Yu ,The End of an Entanglement. \textit{Science} \textbf{316}, 555-557(2007).


\bibitem{Plenio} M. B. Plenio \textit{et al.} Cavity-loss-induced generation of entangled atoms, \pra \textbf{59}.3: 2468 (1999).

\bibitem{Otten} M. Otten,S. K.  Gray, and G. V. Kolmakov,  Optical detection and storage of entanglement in plasmonically coupled quantum-dot qubits. \pra \textbf{99}(3), 032339 (2019).

\bibitem{Kauffman} H. Kaufmann \textit{et al.} Scalable Creation of Long-Lived Multipartite Entanglement. \prl \textbf{119}, 150503 (2017) 

\bibitem{Nourmandipour} A. Nourmandipour \textit{et al.} Entanglement protection of classically driven qubits in a lossy cavity. \textit{Sci. Rep.} \textbf{11}, 16259 (2021).

\bibitem{Bruss} D. Bruss \textit{et al.} Quantum computing with controlled-NOT and few qubits, \textit{Phil.
Trans. R. Soc. Lond. A} {\bf 355},  pp. 2259 (1997).

\bibitem{GHZ1} D.~M. Greenberger, M.~A. Horne, A. Shimony, and A. Zeilinger, Am.
J. Phys. {\bf 58}, 1131 (1990).


\bibitem{GHZ2} D.~M. Greenberger, M.~A. Horne, and A.~Zeilinger, in {\it Bell's
Theorem, Quantum Theory, and Conceptions of the Universe}, edited by M. Kafatos (Springer, Dordrecht, 1989), pp. 6972.




\bibitem{Ekert} A. Ekert and R. Jozsa, \rmp {\bf 68}, 733 (1996).


\bibitem{Shor} P.~W. Shor, \pra {\bf 52}, 2493 (1995).


\bibitem{Steane} A.~M. Steane, Proc. Roy. Soc. Lond. A {\bf 452}, 2551 (1996).


\bibitem{Dicarlo} L. Dicarlo, et al., Nature {\bf 467}, 574
(2010).

\bibitem{Neeley} M. Neeley,  et al.,  Nature {\bf 467}, 570 (2010).

\bibitem{Haffner} H. H\"{a}ffner,  et al., Nature {\bf 438}, 643 (2005).

\bibitem{Neumann} P. Neumann, et al., Science {\bf 323}, 1326 (2009).


\bibitem{Takeda} K. Takeda, A. Noiri, T. Nakajima, J. Yoneda, T. Kobayashi, and
S. Tarucha, Nature Nanotechnology {\bf 16}, 965 (2021).

\bibitem{fidelity} K. Takeda \textit{et al.} Quantum tomography of an entangled three-qubit state in silicon, \textit{Nature Nanotechnology} \textbf{16}, 965 (2021).


\bibitem{exc} M. Kasha, H. R. Rawls, and M. Ashraf El-Bayoumi, The exciton model in molecular spectroscopy, \textit{Pure and Applied Chemistry VIIIth}, \textbf{11}(3-4), 371-392 (1965).



\bibitem{PMF} F. Guinea, M. Katsnelson, and A. Geim, Energy gaps and a zero-field quantum Hall effect in graphene by strain engineering. \textit{Nature Phys} \textbf{6}, 30–33 (2010).


\bibitem{PMF2} S. Zhu, J. A. Stroscio, and T. Li, Programmable extreme pseudomagnetic fields in graphene by a uniaxial stretch. \prl \textbf{115}, 245501 (2015).

\bibitem{PMF3} J. Kim \textit{et al.}, Ultrafast generation of pseudo-magnetic field for valley excitons in WSe2 monolayers. \textit{Science}, \textbf{346} 6214 1205-1208 (2014).

\bibitem{TCHam} M. Tavis, and F. W. Cummings, Exact solution for an N-molecule—radiation-field Hamiltonian, \textit{Phys. Rev.}, \textbf{170}(2), 379 (1968).



\bibitem{BKL} O. L. Berman, R. Y. Kezerashvili, Y. E. Lozovik, amd K. Ziegler, Strain-induced quantum Hall phenomena of excitons in graphene, \textit{Sci Rep} \textbf{12}, 2950 (2022). 

\bibitem{2qb} G. P. Martins, O. L. Berman, G. Gumbs, and Y. E. Lozovik Quantum entanglement between excitons in two-dimensional materials, \prb \textbf{106}, 104304 (2022).

\bibitem{law} C. K. Law, Effective Hamiltonian for the radiation in a cavity with a moving mirror and a time-varying dielectric medium, \pra \textbf{49} 433 (1994).

\bibitem{optics} C. Gerry, and P. Knight, \textit{Introductory Quantum Optics}  (Cambridge University Press, New
York, 2004).

\bibitem{albert} F. Albert \textit{et al.} Microcavity controlled coupling of excitonic qubits Nat. Comm. \textbf{4} 1473 (2013).

\bibitem{Levy} N. Levy \textit{et al.} Strain-Induced Pseudo–Magnetic Fields Greater Than 300 Tesla in Graphene Nanobubbles, \textit{Science} \textbf{329},544-547(2010). 

\bibitem{cav} Q. Zhang \textit{et al.} Collective non-perturbative coupling of 2D electrons with high-quality-factor terahertz cavity photons, \textit{Nature Physics} \textbf{12}.11 : 1005-1011 (2016).

\bibitem{laser} A. Khalatpour \textit{et al.} High-power portable terahertz laser systems, \textit{Nat. Photonics} \textbf{15}, 16–20 (2021).

\bibitem{cav_lifetime} K. Kristinsson, O. Kyriienko, and I. A. Shelykh, Terahertz laser based on dipolaritons, \pra \textbf{89}.2 023836 (2014).


\bibitem{Fogler1} M.~M. Fogler, L.~V. Butov, and K.~S. Novoselov, Nat. Commun. {\bf 5},
4555 (2014).

\bibitem{Fogler2} E.~V. Calman, M.~M. Fogler, L.~V. Butov, S. Hu, A. Mishchenko, and
A.~K. Geim, Nat. Commun. {\bf 9},  1895 (2018).


\bibitem{Loz1} I.V.Lerner, Yu.E.Lozovik, Mott exciton in a quasi-two-dimensional semiconductor in a strong magnetic field, \textit{JETP} \textbf{51}, N3, 588-592(1980).

\bibitem{Loz2} I.V.Lerner, Yu.E.Lozovik, Two - dimensional electron - hole systems in a strong magnetic field as the almost ideal excitons gas, \textit{JETP} \textbf{53}, N4, 763-770(1981).

\bibitem{Loz3} A.B.Dzyubenko, Yu.E.Lozovik, Quasi-two-dimensional electron - hole pairs condensate in the strong magnetic fields, \textit{Solid State Phys.} \textbf{26}, 5, 938(1984).

\bibitem{Loz4} A.B.Dzyubenko, Yu.E.Lozovik, Symmetry of Hamiltonians of quantum two - component systems: condensate of composite particles as an exact eigenstate, \textit{Journ. Phys. A} \textbf{24}, 415-424(1991).

\bibitem{lim1} L. T. Shen \textit{et al.} Steady-state entanglement for distant atoms by dissipation in coupled cavities, \pra \textbf{84}, 064302 (2011).

\bibitem{lim2} Z. Ficek, and S. Natali  Diffraction effects in
entanglement of two distant atoms. \textit{Journ. of Phys.:
Conference Series}, \textbf{84}. 12007 (2007).

\end{thebibliography}
\end{document}